\documentclass{aa}
\usepackage{graphicx}
\usepackage{epsf,epsfig}

\begin{document}

   \title{Bispectrum speckle interferometry of the B[e] star MWC\,349A}

\author{K.-H.\,Hofmann\inst1, Y.\,Balega\inst2,
N.~R.\,Ikhsanov\inst1\fnmsep\inst3,
A.~S.\,Miroshnichenko\inst1\fnmsep\inst3\fnmsep\inst4,
G.\,Weigelt\inst1}

  \offprints{K.-H.\,Hofmann \\ \email{khh@mpifr-bonn.mpg.de}}

   \institute{Max-Planck-Institut f\"ur Radioastronomie, Auf dem
              H\"ugel 69, D-53121 Bonn, Germany
              \and
              Special Astrophysical Observatory, Nizhnij Arkhyz,
              Zelenchuk district, 369167, Karachai-Chirkassian Republic, Russia
              \and
              Central Astronomical Observatory RAS at Pulkovo,
              Pulkovo 65/1, 196140, St.\,Petersburg, Russia
              \and
              Department of Physics and Astronomy, University of Toledo, Toledo,
              OH 43606, U.S.A.}

  \date{Received ; accepted }

\authorrunning{K.-H. Hofmann et al.}

\abstract{We present the results of bispectrum speckle
interferometry of the B[e] star MWC\,349A obtained with the SAO
6\,m telescope. Our diffraction-limited $J$-, $H$-, and $K$-band
images (resolutions 43--74\,mas) suggest the star is surrounded by
a circumstellar disk seen almost edge-on. The observed visibility
shape is consistent with a two-component elliptical disk model,
probably corresponding to the gaseous and dusty components of the
disk. We show that the classification of the object as a
pre-main-sequence star or a young planetary nebula is problematic.
An analysis of the uncertainties in the basic parameter
determination lead us to the conclusion that MWC\,349A is probably
either a B[e] supergiant or a binary system, in which the
B[e]-companion dominates the observed properties.
\keywords{techniques: image processing -- stars: emission-line --
stars: individual: MWC\,349A}}

   \maketitle

   \section{Introduction}

\object{MWC\,349} consists of two apparently close (the angular
distance is 2.4\,arcsec) objects: \object{MWC\,349A}, which
exhibits a strong emission-line spectrum and IR-excess, and
\object{MWC\,349B}, a weak emission-line source which is
$\sim$40\% fainter than MWC\,349A in the optical region (Cohen et
al. \cite{cbdw85}, hereafter C85). C85 argued that MWC\,349A and
MWC\,349B are a physical pair, as they are connected by an arc of
emitting matter, and estimated the distance (1.2\,kpc) toward it
based on the spectral type and luminosity of MWC\,349B (B0\,{\sc
iii} and M$_{V}=-$5.0\,mag, respectively). However, Meyer et al.
(\cite{mnh02}) found that the A and B components have a different
level of interstellar polarization, suggesting that they are not
connected to each other and that MWC\,349A could be a member of
Cyg\,OB2. In this paper we concentrate on MWC\,349A and use
MWC\,349B only as a reference star for our observations.

MWC\,349A is a peculiar stellar object with one of the strongest
emission-line spectra ever observed. It is located in the southern
part of the \object{Cyg OB2} association, a heavily reddened
stellar cluster at a distance of 1.7\,kpc (e.g., Kn\"odlseder
\cite{knodl00}). The object does not show any photospheric lines
in its spectrum (Andrillat, Jaschek, \& Jaschek \cite{ajj96}),
although the presence of strong He\,{\sc i} emission lines
indicates a high temperature of the underlying source. In most of
the studies devoted to the nature of MWC\,349A its spectral type
is estimated as late O (e.g., Hartmann, Jaffe, \& Huchra
\cite{hjh80}). The co-existence of forbidden emission lines in the
optical spectrum and a near-IR excess brought MWC\,349 into the
group of B[e] stars (Allen \& Swings \cite{as76}). On the other
hand, MWC\,349A is considered to be a pre-main-sequence object
because of the surrounding nebula, which is formed by an
optically-thick bipolar outflow seen at radio wavelengths (Olnon
\cite{o75}). The bipolar structure of the nebula suggests that the
stellar core is surrounded by a disk, whose presence is inferred
from a high level of polarization (Elvius \cite{elv74}, Zickgraf
\& Schulte-Ladbeck \cite{zsl89}, Yudin \cite{y96}; Meyer,
Nordsieck, \& Hoffman \cite{mnh02}) and double-peaked
emission-line profiles (Hartmann et al. \cite{hjh80}). MWC\,349A
also shows maser and laser line emission in the IR and radio
spectral regions, which makes it rather unique among similar
sources (see Gordon et al. \cite{g01} for a recent update).

Since MWC\,349A is a distant object, observations with high
spatial resolution are crucial for revealing its nature. So far
results of only a few high-resolution near-IR observations have
been published. Leinert (\cite{l86}) marginally resolved the IR
source and found it elongated with a gaussian FWHM of the
brightness distribution of $85\pm19$\,mas in the east-west
direction at L$^{\prime}$ and $38\pm18$\,mas in the north-south
direction at K. The highest resolution measurements were presented
by Danchi, Tuthill, \& Monnier (\cite{dtm01}). These authors
showed that the IR source can be fitted by uniform ellipses with
major axes of $36\pm2$\,mas at 1.65\,$\,\mu$m, 47$\pm$2\,mas at
2.25\,$\,\mu$m and $62\pm1$\,mas at 3.08\,$\,\mu$m, axial ratios
of $\sim$0.5, and a position angle of $100^{\degr} \pm 3^{\degr}$.
Under a flat disk approximation, Danchi et al. estimated the
inclination angle of the disk plane to the line of sight to be
$\sim21^{\degr}$. They also found the observed disk sizes at the
above wavelengths consistent with the theoretical predictions for
the inner and outer dimensions of photoevaporating accretion disks
around young stars reported by Hollenbach et al. (\cite{hjls94}).

To verify the results by Danchi et al. (\cite{dtm01}) and to
extend the spatial information toward shorter wavelengths, we
obtained new speckle interferometric observations of MWC\,349A in
$J$, $H$, and $K$ broadband filters and a narrowband filter at
2.09 $\mu$m. Our results are presented in Sect.~\ref{obsres}.
Additionally, we review the existing information about the object,
as its nature and evolutionary state remain uncertain. Most of the
investigators focus on details of the circumstellar (CS)
structures (mainly on the disk) in efforts to interpret the
observed features, while properties of the illuminating source are
only vaguely known. In Sect.~\ref{nature} we re-estimate physical
parameters of the object and examine existing hypotheses on its
nature. In Sect.~\ref{discuss} we consider the reliability of the
adopted parameters and summarize our findings in
Sect.~\ref{conclus}.

   \begin{figure*}
  \begin{center}
\epsfxsize=50mm \mbox{\epsffile{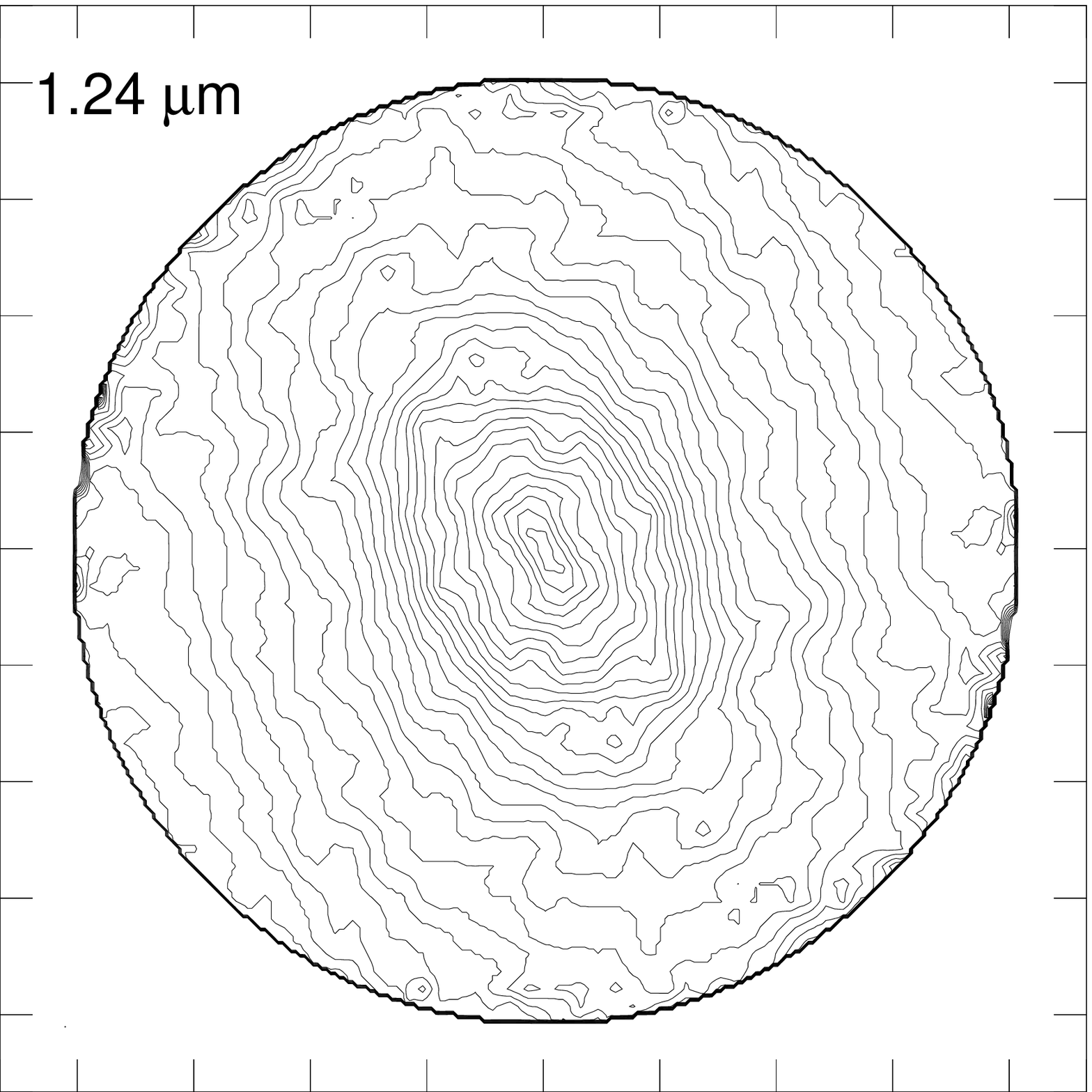}}\hspace{4mm}
\epsfxsize=56mm \mbox{\epsffile{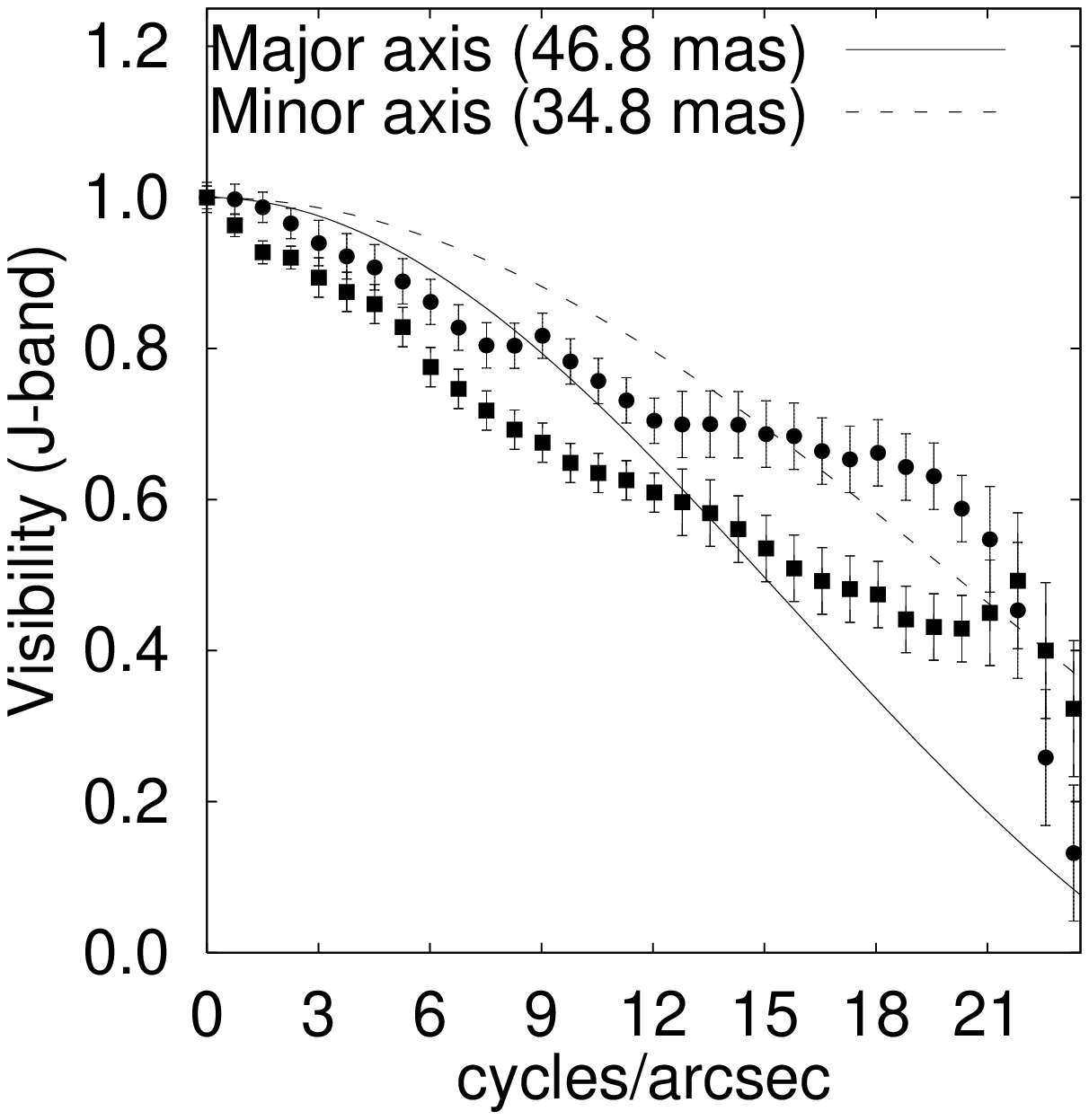}}
\epsfxsize=50mm \mbox{\epsffile{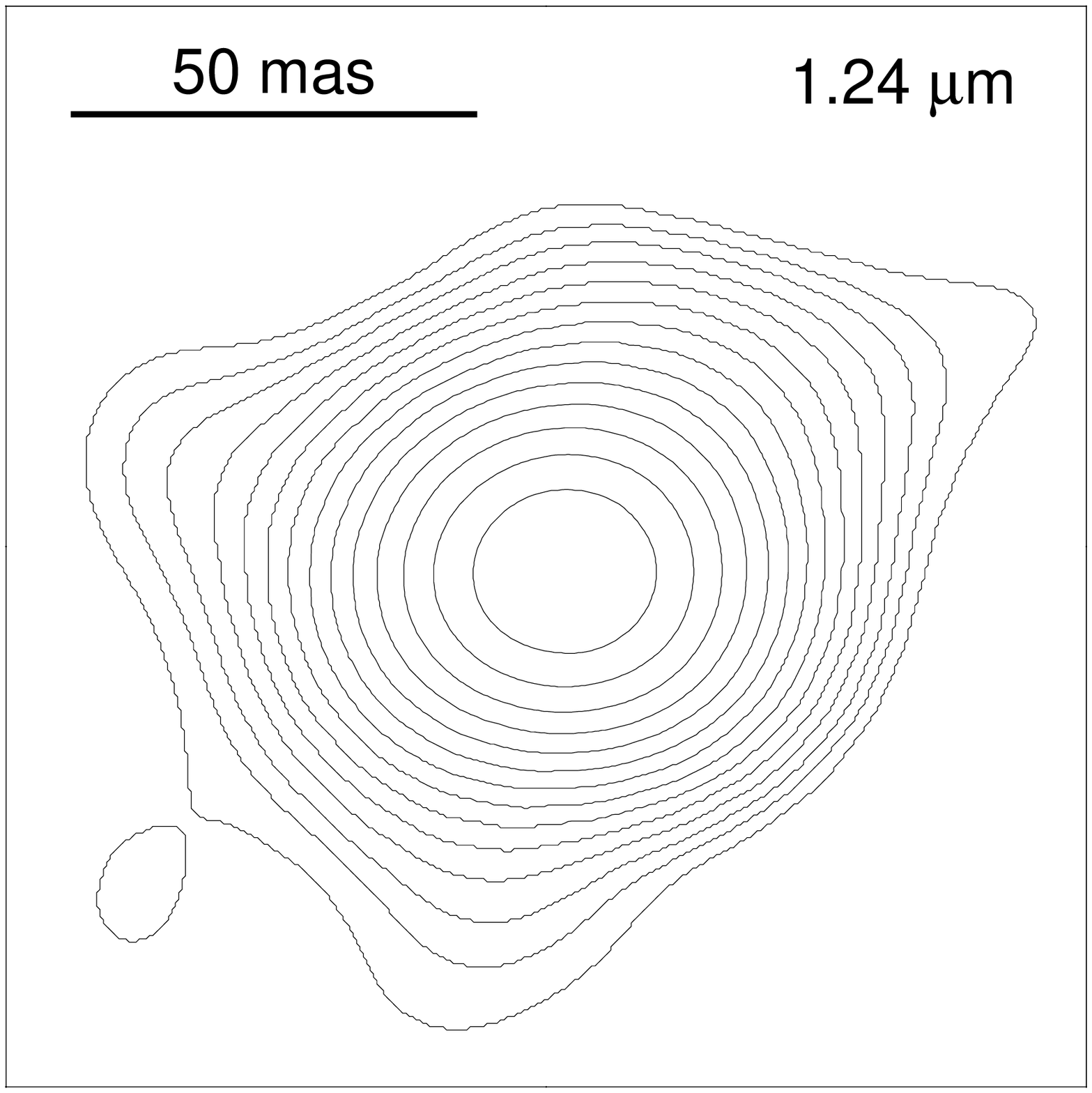}}\\[4mm]
\vspace{0.5cm}

\epsfxsize=50mm \mbox{\epsffile{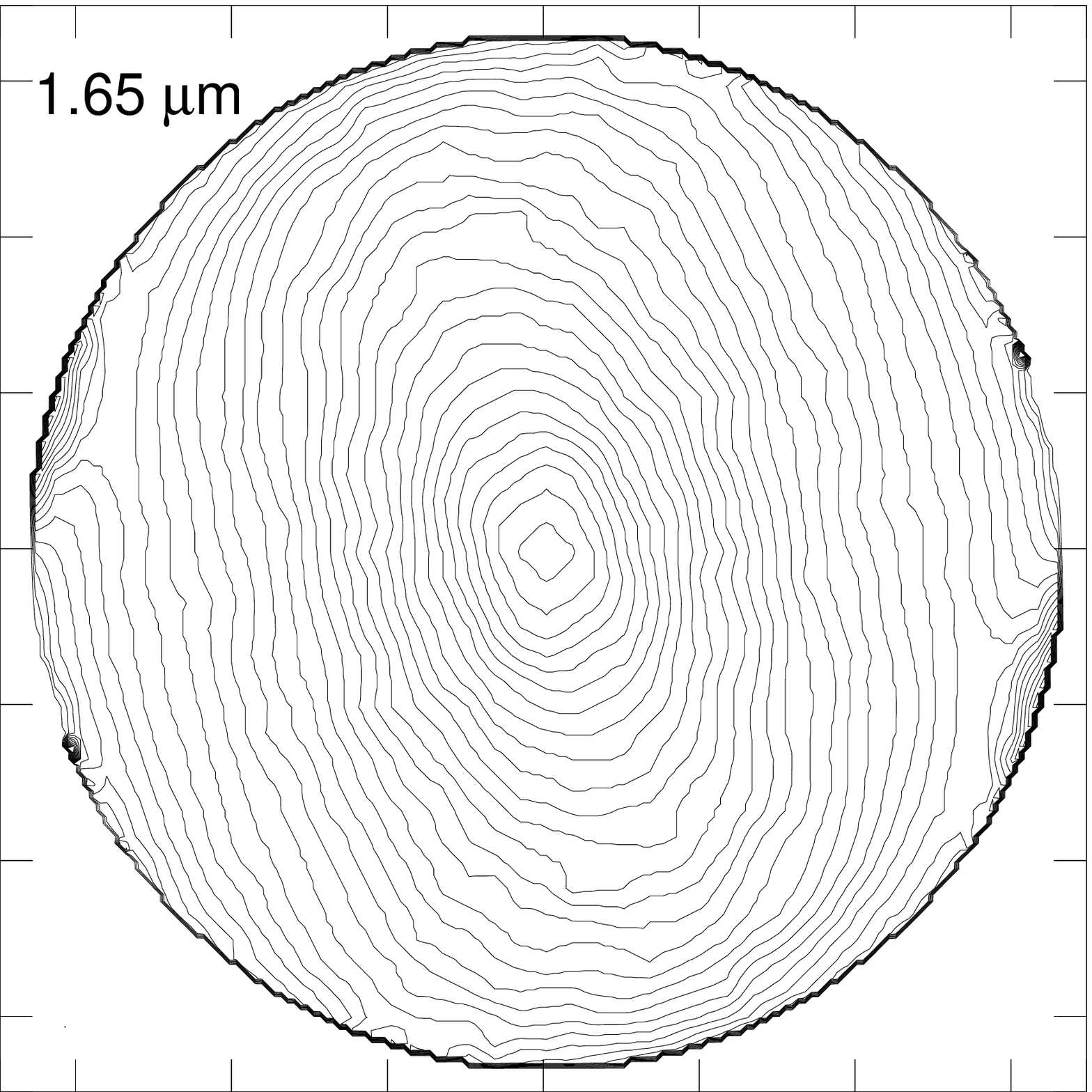}}\hspace{4mm}
\epsfxsize=56mm \mbox{\epsffile{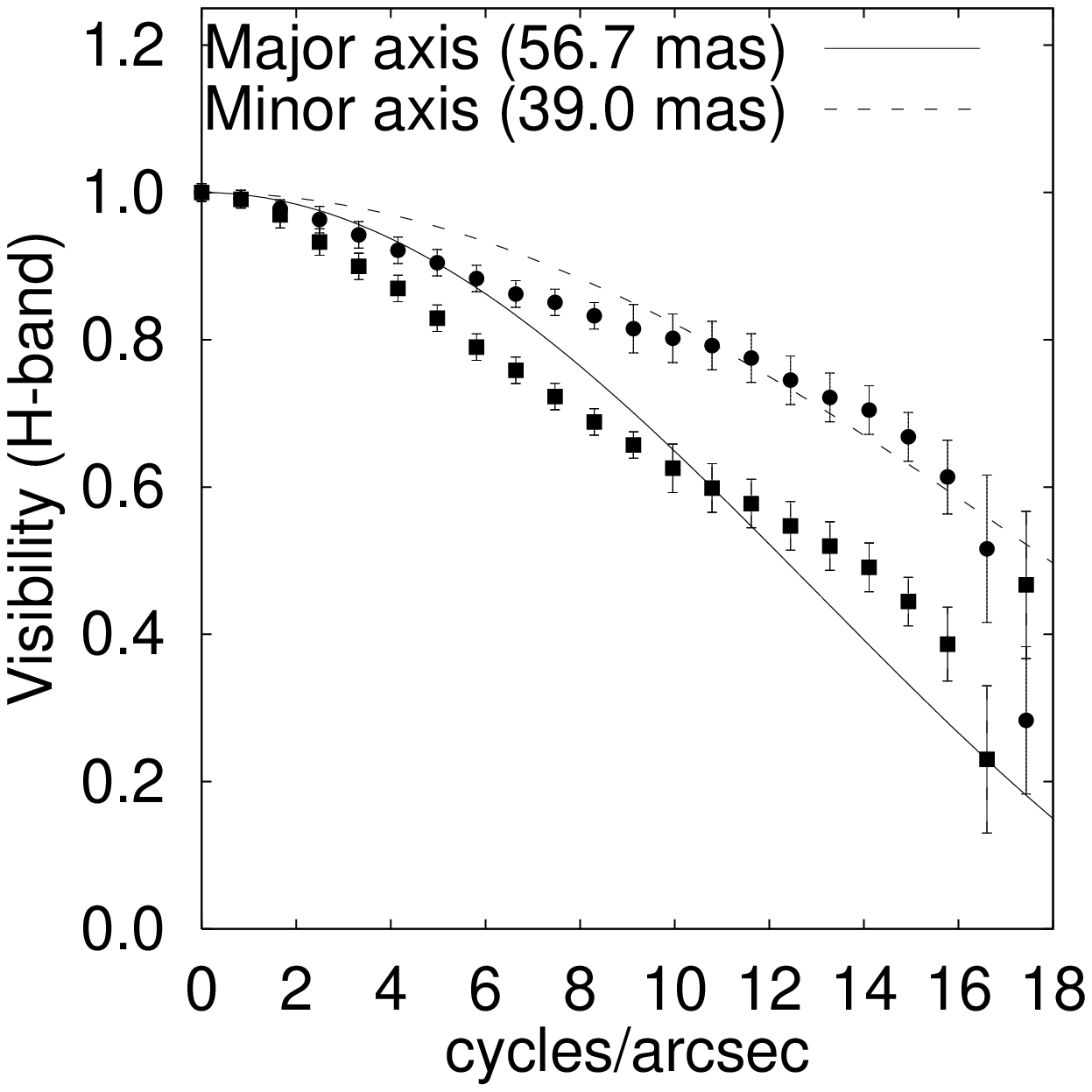}}
\epsfxsize=50mm \mbox{\epsffile{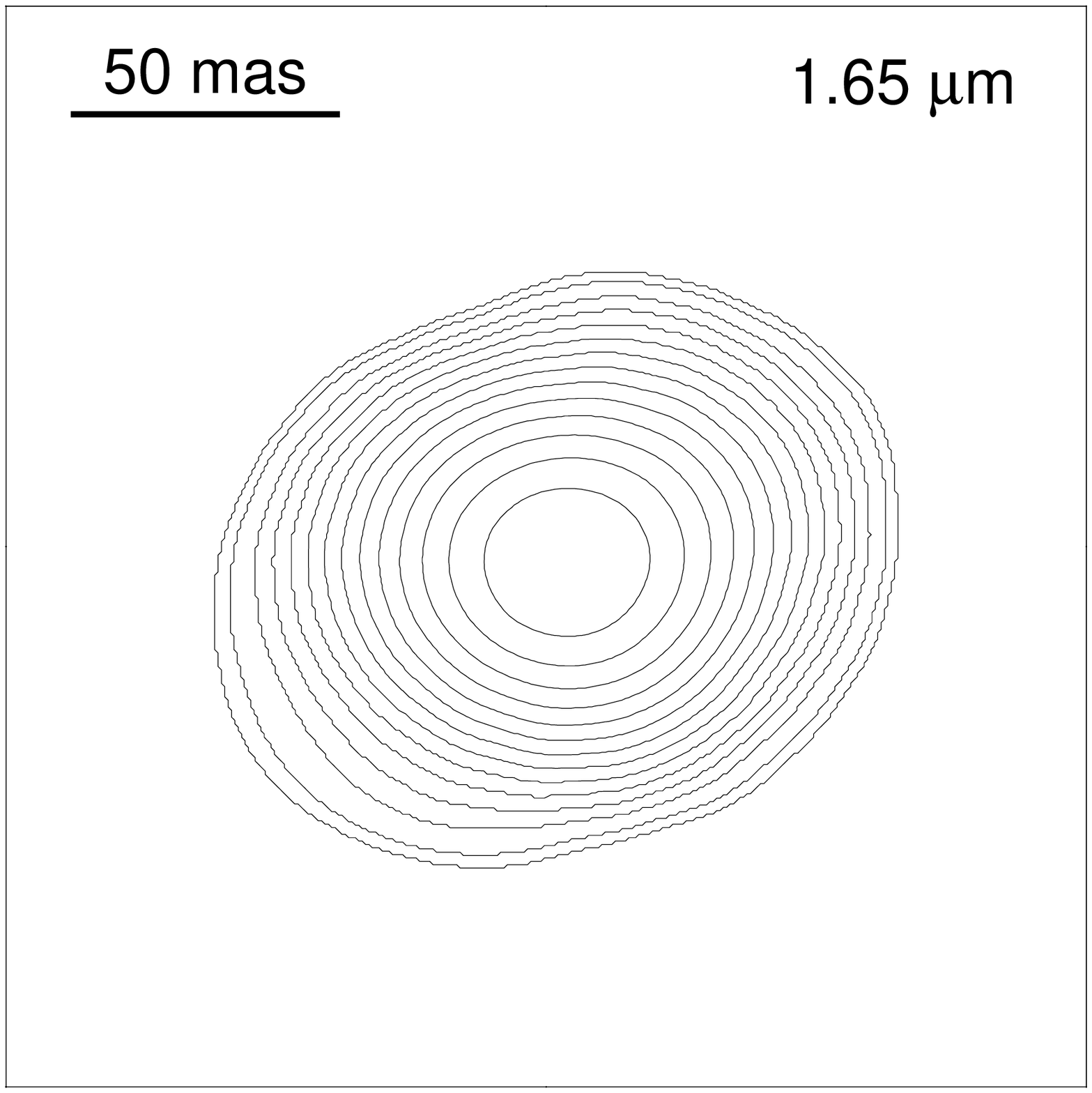}}\\[4mm]
\vspace{0.5cm}

\epsfxsize=50mm \mbox{\epsffile{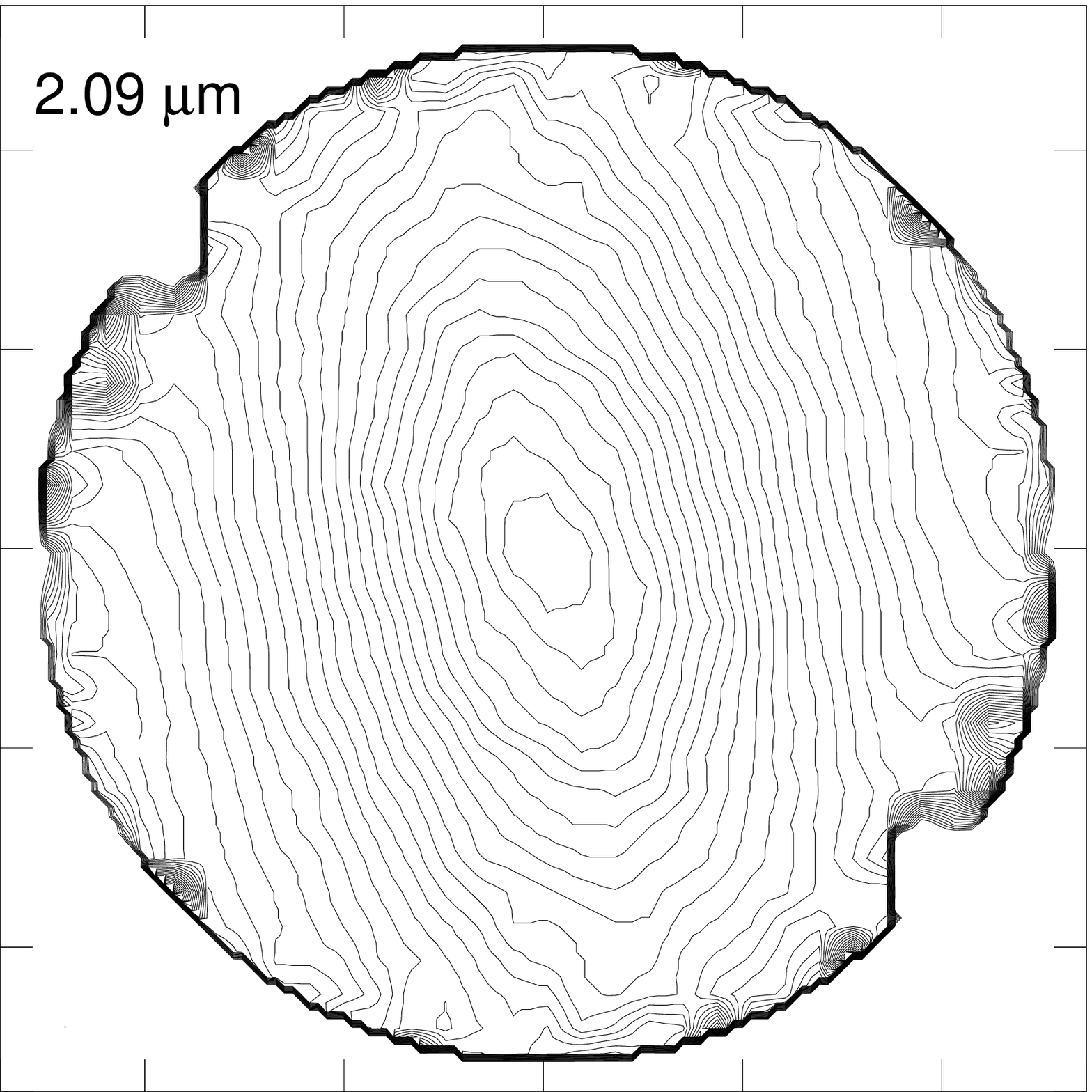}}\hspace{4mm}
\epsfxsize=56mm \mbox{\epsffile{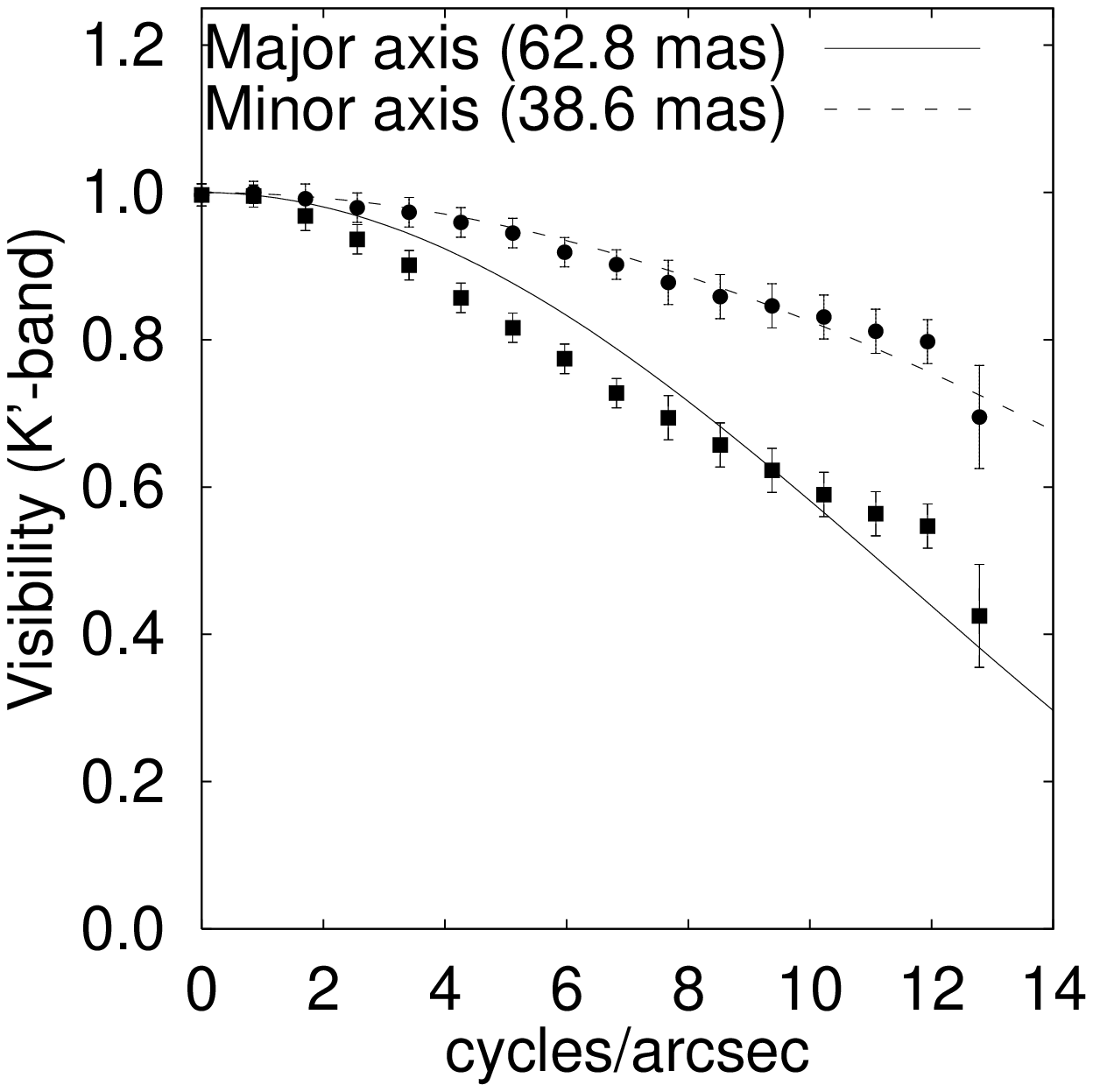}}
\epsfxsize=50mm \mbox{\epsffile{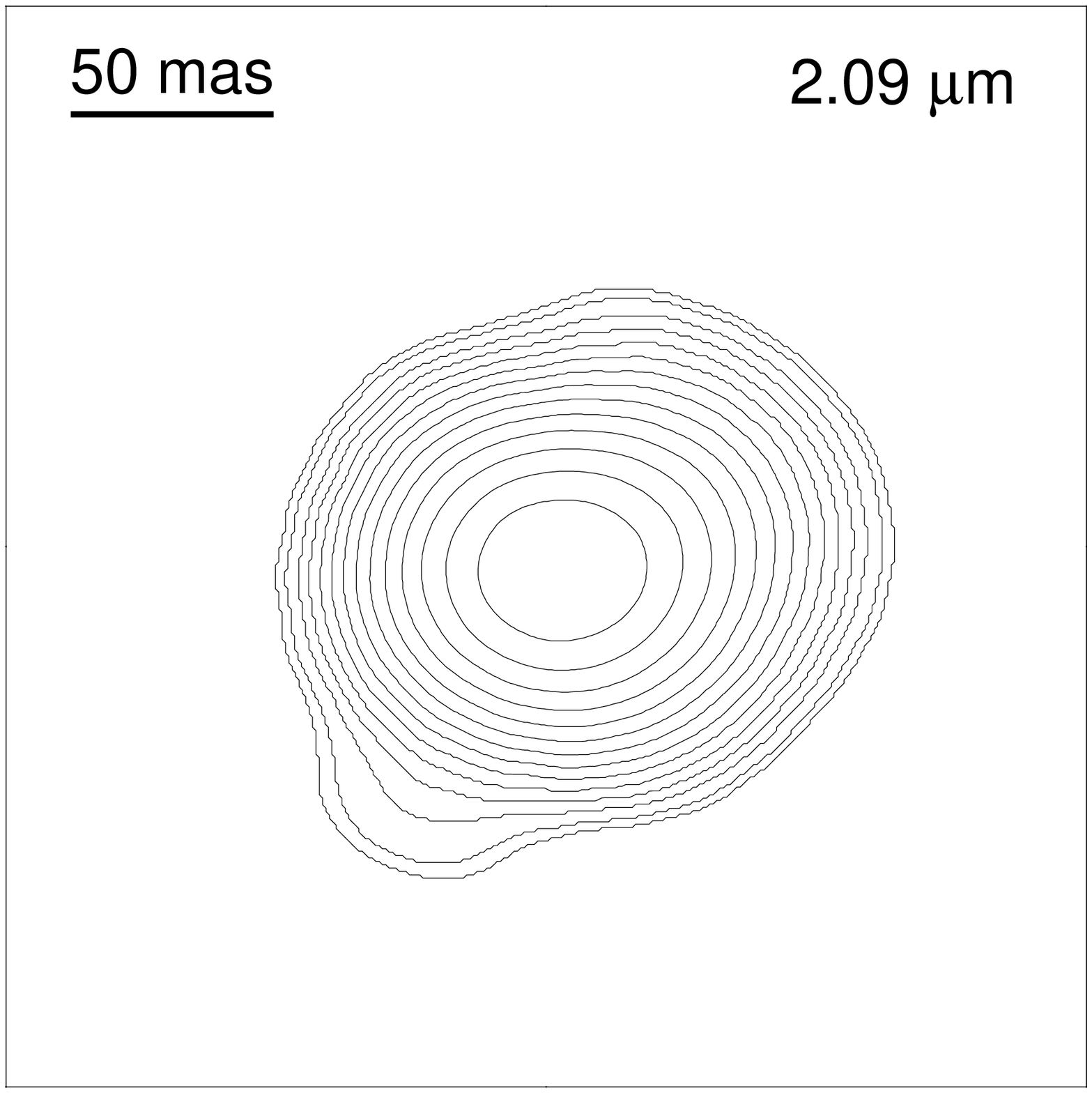}}\\[4mm]
\vspace{0.5cm}

\epsfxsize=50mm \mbox{\epsffile{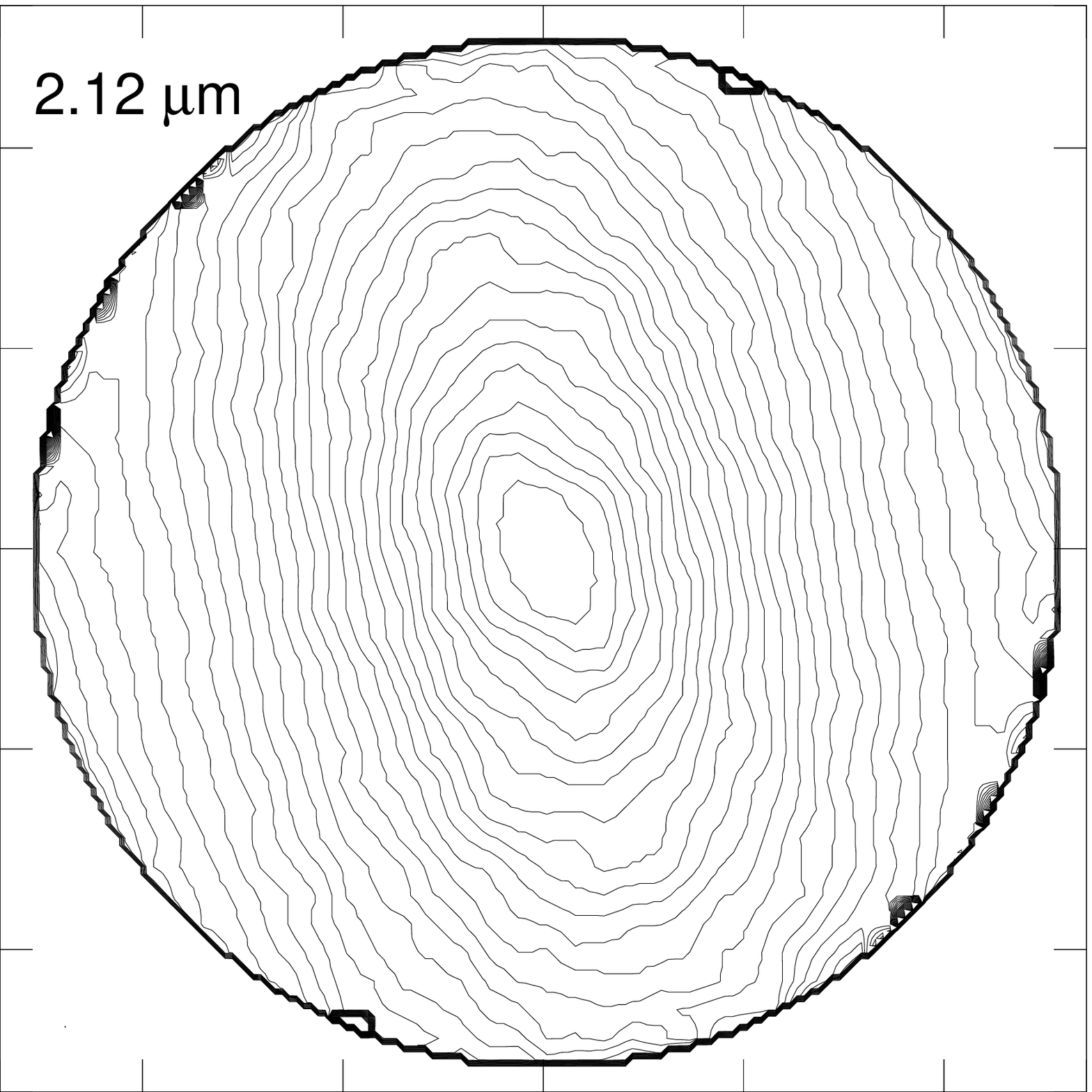}}\hspace{4mm}
\epsfxsize=56mm \mbox{\epsffile{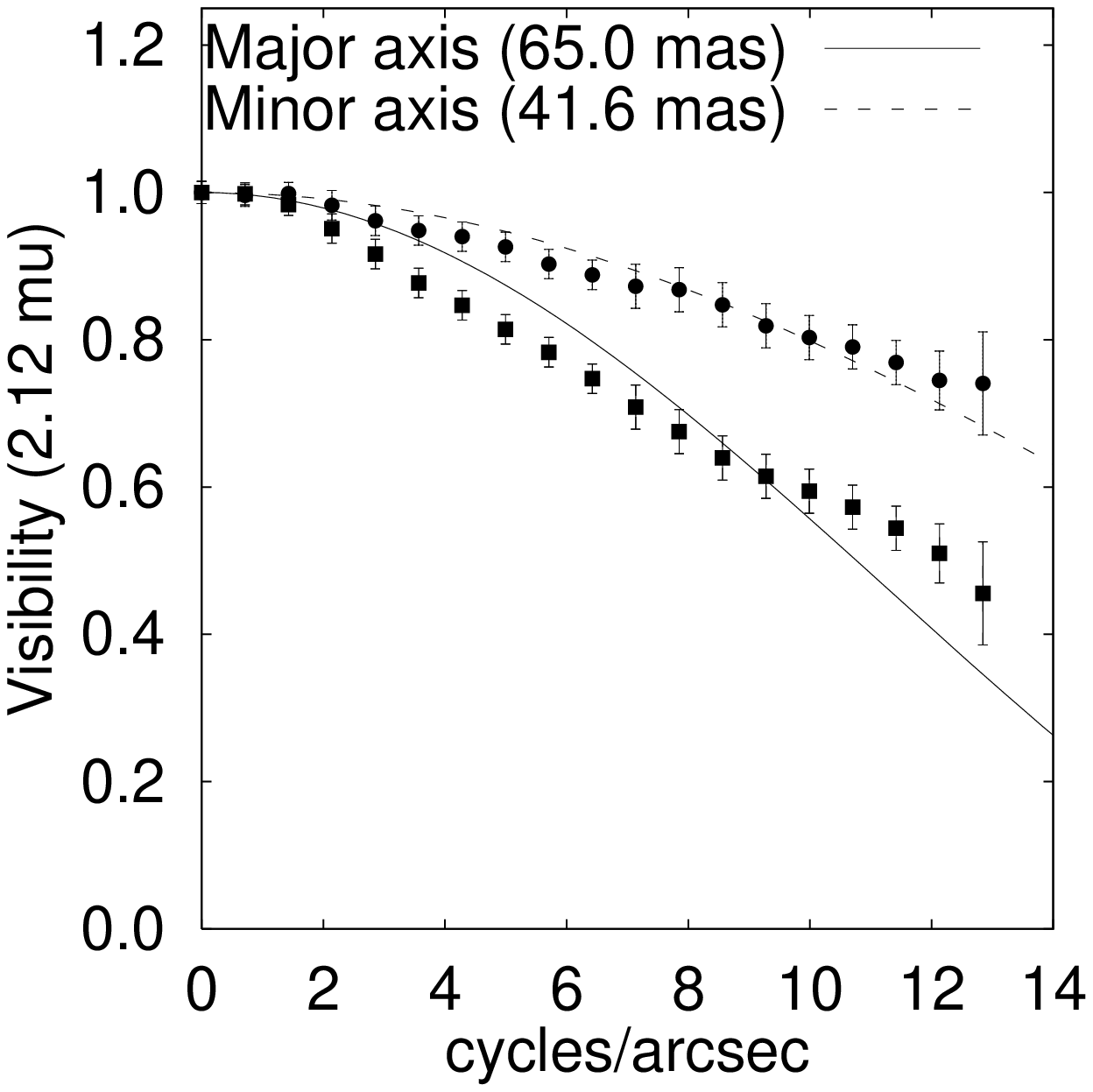}}
\epsfxsize=50mm \mbox{\epsffile{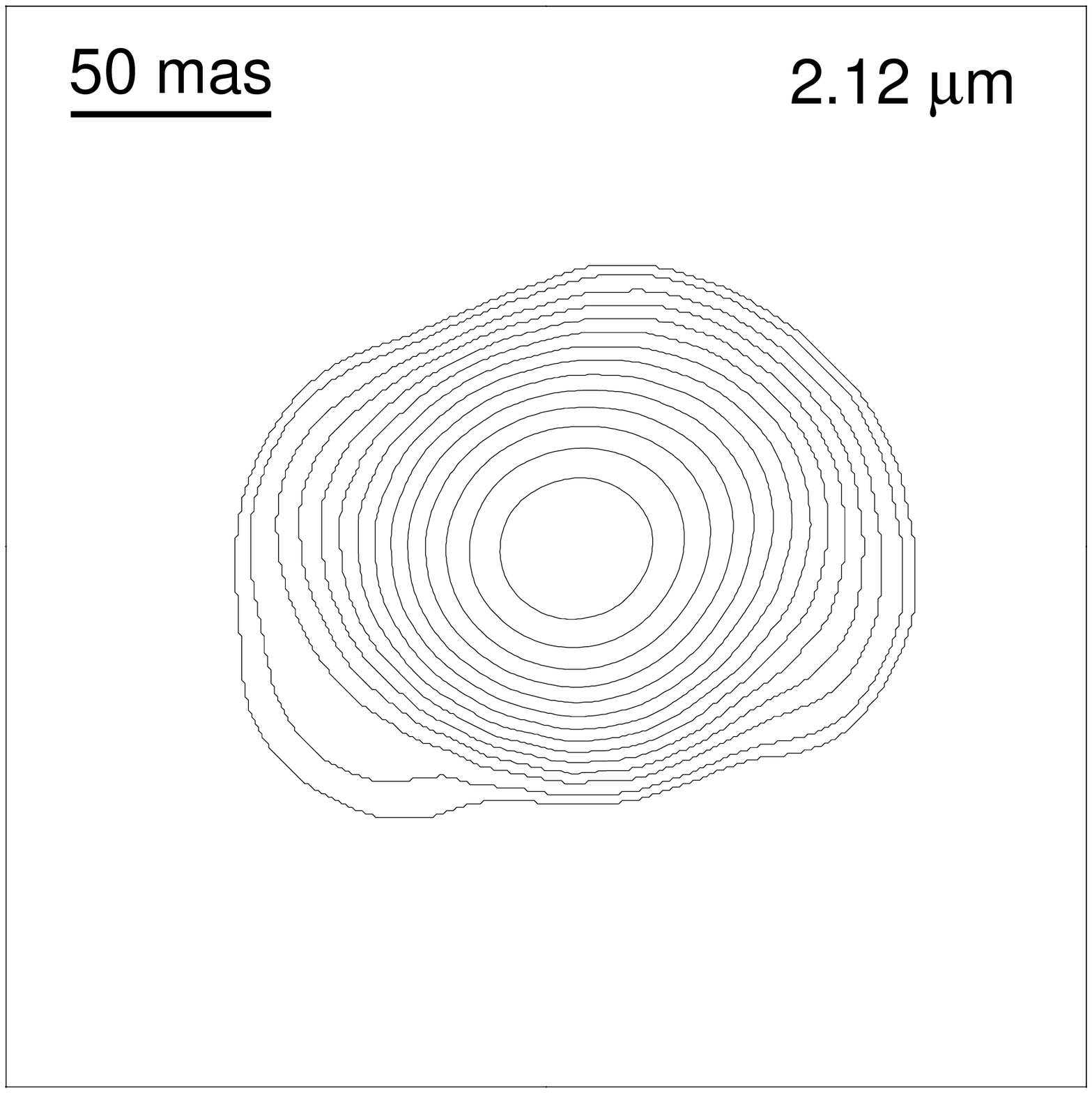}}\\[4mm]
\end{center}
\caption{Left column: two-dimensional visibilities. Middle column:
cuts through the long and short axis of the visibilities and the
visibilities of the best-fitting two-dimensional elliptical
uniform-disk models fitted up to the telescope cut-off frequency.
Right column: images of MWC\,349A reconstructed by bispectrum
speckle interferometry. The contours are plotted in steps of
0.3\,mag (from 0.3\,mag to 4.2\,mag of peak intensity).}
   \label{f1}
 \end{figure*}

   \begin{figure*}
  \begin{center}
\epsfxsize=66mm \mbox{\epsffile{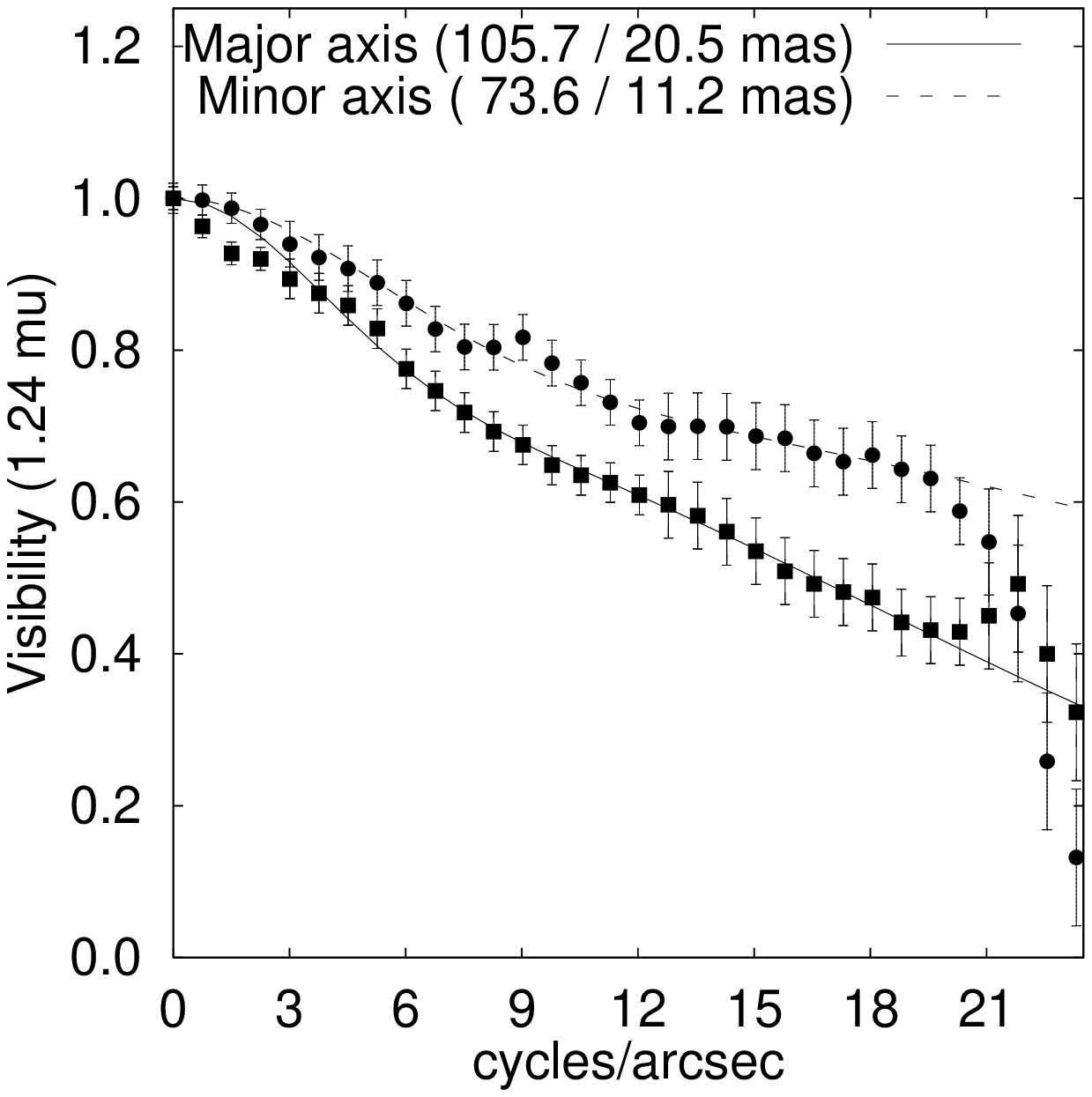}}\hspace{4mm}
\epsfxsize=66mm \mbox{\epsffile{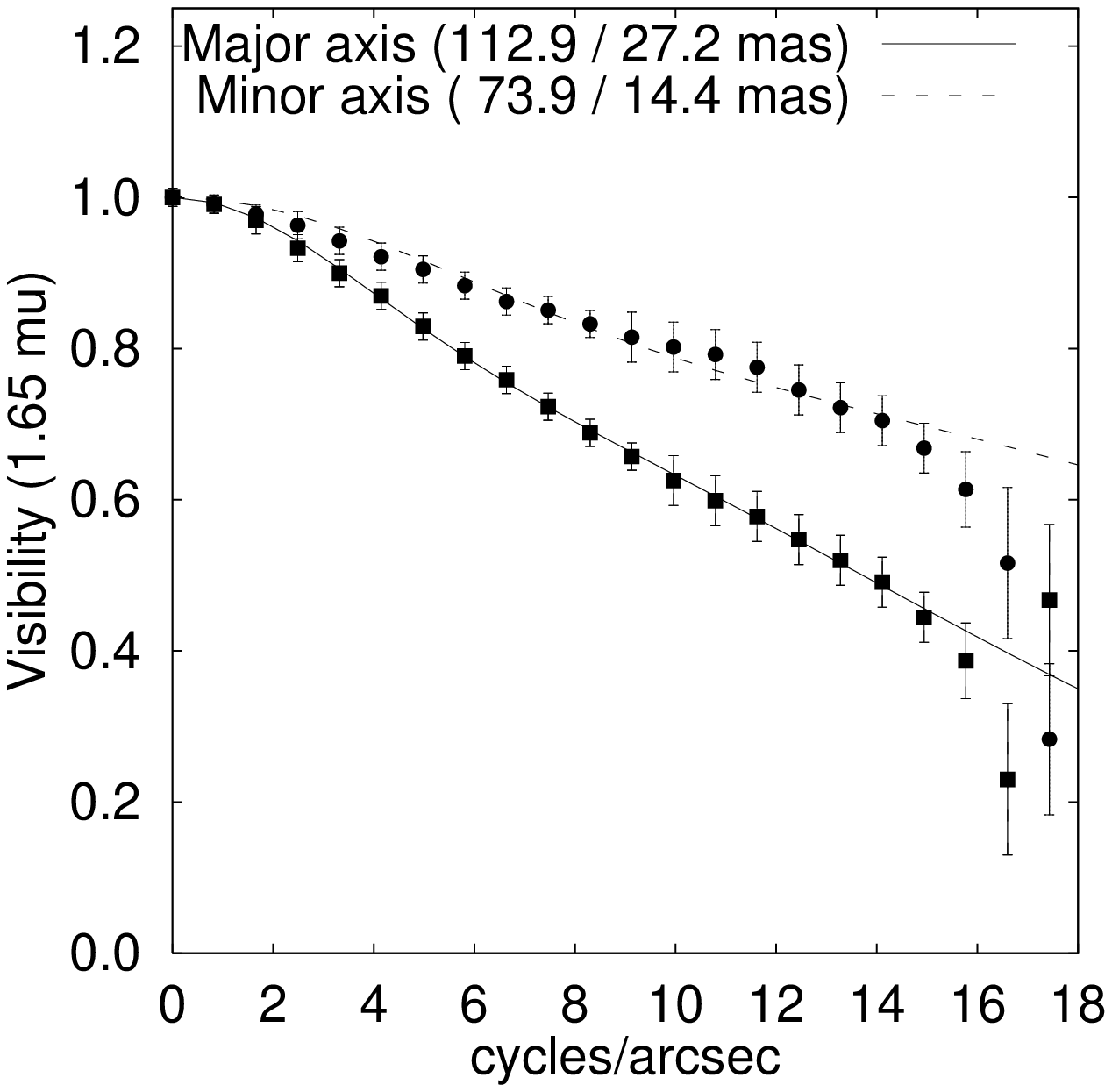}}\\[8mm]
\epsfxsize=66mm \mbox{\epsffile{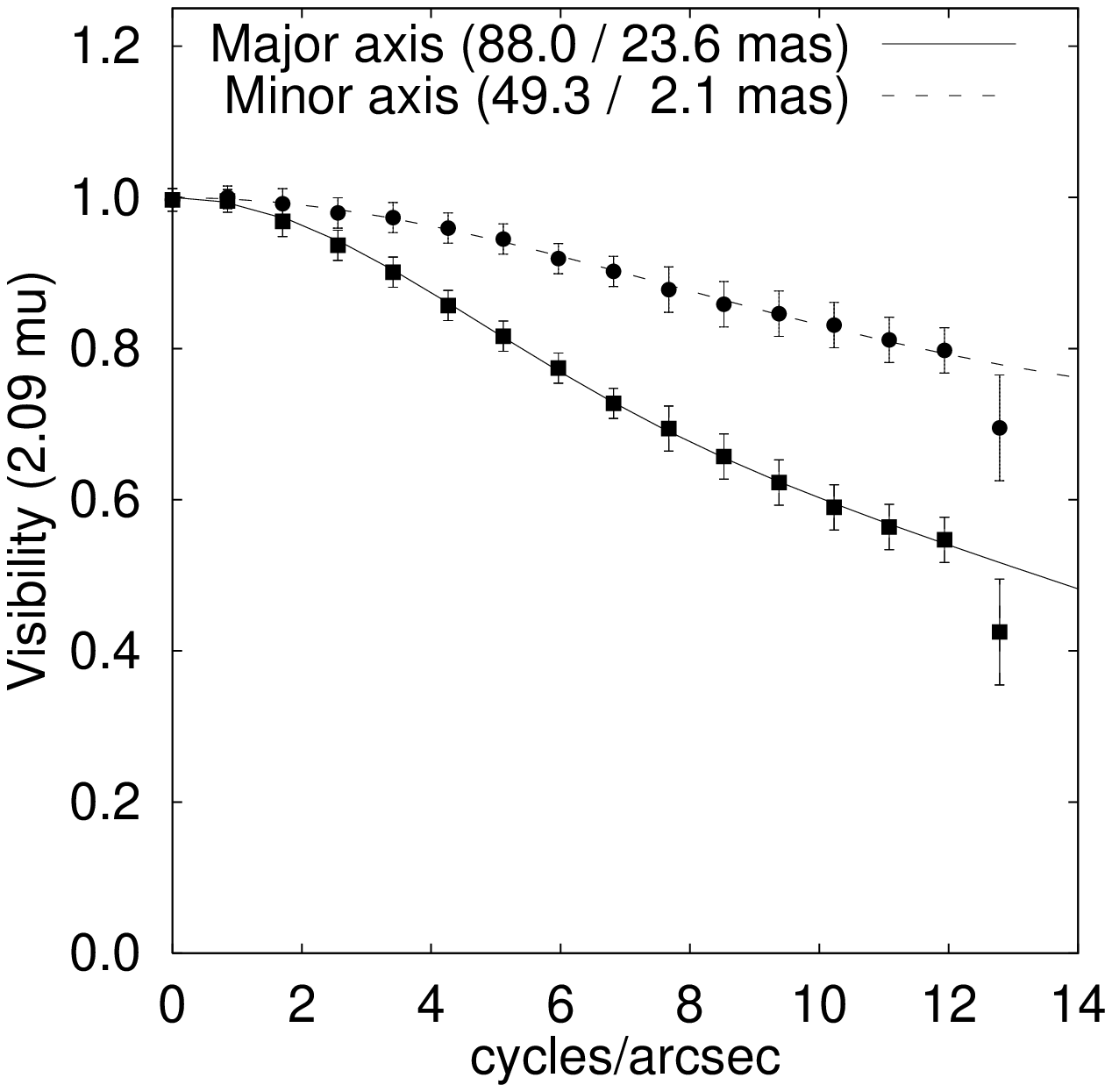}}\hspace{4mm}
\epsfxsize=66mm \mbox{\epsffile{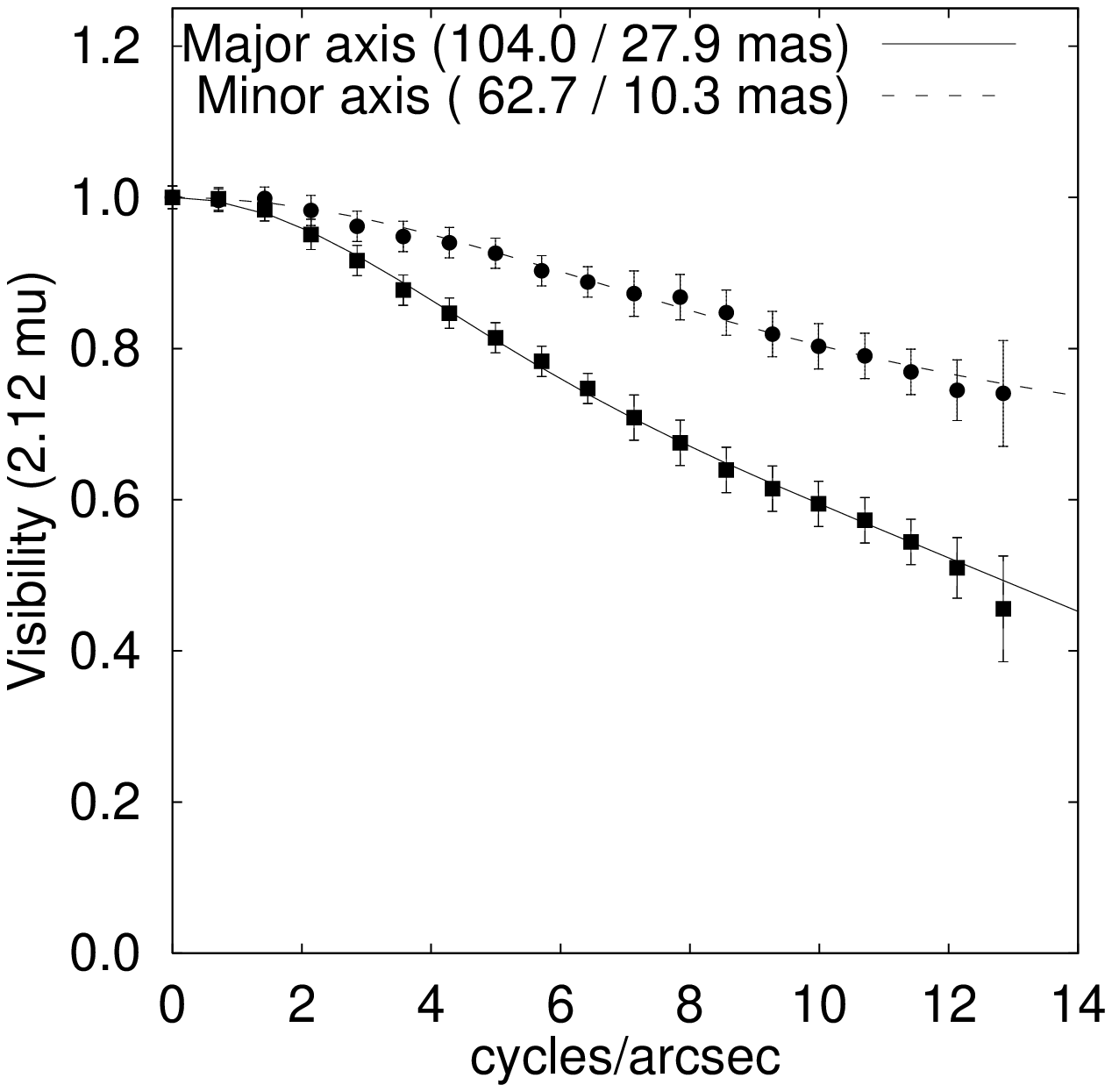}}\\[8mm]
\end{center}
\caption{The visibilities of the best-fitting two-component
visibility models consisting of two Gaussian functions fitted to
the cuts through the short and long axes of the measured
visibilities (corresponding to the major and minor axes of the
Gaussian disk in image space). The fit range chosen is up to the
telescope cut-off frequency. The numbers in parenthesis are the
FWHM diameters  of the large/small component of the two-component
model.}
   \label{f2}
 \end{figure*}

   \section{Observations}\label{obsres}

The speckle interferograms of MWC\,349A were obtained with the
Russian 6\,m telescope of the Special Astrophysical Observatory
(SAO) on October 28, 2001 and November 1, 2001. The data were
taken with our HAWAII speckle camera through interference filters
with center wavelengths/bandwidths of 1.24$\mu$m/0.14$\mu$m,
1.65$\mu$m/0.32$\mu$m, 2.12$\mu$m/0.21$\mu$m and
2.09$\mu$m/0.02$\mu$m. The simultaneously recorded speckle
interferograms of the unresolved star MWC\,349B were used for the
compensation of the atmospheric speckle transfer function. The
observational parameters are summarized in Table~\ref{t1}.

\begin{table}[t]
 \caption[]{Observational parameters. $\lambda_c$ is the central
 wavelength and $\Delta\lambda$ the FWHM bandwidth of the filters.
 $T$ denotes the exposure time per frame and $N$ is the number of
 frames. $S$ denotes the seeing (FWHM) and $p$ is the pixel size.}
   \label{t1}
\begin{tabular}{cccccr}
  \noalign{\smallskip}
  \hline
  \noalign{\smallskip}
  $\lambda_c$ & $\Delta\lambda$ & $T$ & $N$ & $S$ & $p$ \\
 $[\mu$m]    & [$\mu$m]        & [ms]        &             & ['']   & [mas] \\
  \noalign{\smallskip}
  \hline
  \noalign{\smallskip}
1.24 & 0.14 & 120 & 695 & 0.7 & 13.30 \\
1.65 & 0.32 &  80 & 383 & 0.9 & 20.08 \\
2.12 & 0.21 &  30 & 855 & 1.2 & 26.95 \\
2.09 & 0.02 & 160 & 820 & 0.7 & 26.66 \\
  \noalign{\smallskip}
  \hline
\end{tabular}
\newline
\end{table}

\begin{table}[t]
 \caption[]{Results of two-dimensional elliptical uniform-disk fits to
   the visibilities of MWC\,349A for a fit range up to the telescope
   cut-off frequency (see discussion on the fit range in the text).}
   \label{t2}
\begin{tabular}{cccc}
  \noalign{\smallskip}
  \hline
  \noalign{\smallskip}
Filter   & Major Axis      & Axial Ratio & Position Angle \\
$[\mu$m] & [diameter, mas] &             & [deg] \\
  \noalign{\smallskip}
  \hline
  \noalign{\smallskip}
1.24/0.14 & $47.5\pm 2.7$ & $0.75\pm 0.04$ & $109.7\pm 5.0$ \\
1.65/0.32 & $56.2\pm 1.8$ & $0.70\pm 0.08$ & $100.9\pm 2.0$ \\
2.12/0.21 & $65.0\pm 2.7$ & $0.63\pm 0.04$ & $101.4\pm 2.0$ \\
2.09/0.02 & $62.5\pm 2.7$ & $0.62\pm 0.04$ & $105.2\pm 2.0$ \\
  \noalign{\smallskip}
  \hline
\end{tabular}
\newline
\end{table}


\begin{table*}[t]
 \caption[]{Two-component fit diameters: Results of one-dimensional fits
 of two-component models
 consisting of two centered Gaussian functions to the cuts through
 the short and long axes of the MWC\,349A visibilities.}
   \label{t3}
\begin{tabular}{cccccc}
  \noalign{\smallskip}
  \hline
  \noalign{\smallskip}
Filter   & FWHM [mas]:      & FWHM [mas]:       & FWHM [mas]:     & FWHM [mas]:     & Brightness Ratio  \\
$[\mu$m] & Large Component  & Large Component   & Small Component & Small Component & Large/Small Component \\
         & Major Axis       & Minor Axis        & Major Axis      & Minor Axis      &  \\
  \noalign{\smallskip}
  \hline
  \noalign{\smallskip}
1.24/0.14 & $118.7\pm 19.6$ & $73.6\pm 14.0$    & $21.8\pm 1.0$   & $12.4\pm 2.0$ & $0.32\pm 0.03$ \\
1.65/0.32 & $115.7\pm  7.6$ & $68.8\pm  7.4$    & $26.9\pm 3.0$   & $15.4\pm 2.3$ & $0.22\pm 0.02$ \\
2.12/0.21 & $ 95.8\pm 11.7$ & $58.1\pm  7.3$    & $25.1\pm 4.6$   & $9.7\pm 6.1 $ & $0.38\pm 0.16$ \\
2.09/0.02 & $ 93.7\pm  8.6$ & $49.4\pm  7.0$    & $25.6\pm 3.1$   & $13.5\pm 8.5$ & $0.34\pm 0.10$ \\
  \noalign{\smallskip}
  \hline
\end{tabular}
\newline
\end{table*}


Diffraction-limited images of MWC\,349A were reconstructed from
the speckle interferograms using the bispectrum speckle
interferometry method (Weigelt \cite{w77}; Lohmann et al.
\cite{lww83}; Hofmann \& Weigelt \cite{Hof86}) and the building
block method (Hofmann \& Weigelt \cite{Hof93}). The modulus of the
object Fourier transform (visibility function) was determined with
the speckle interferometry method (Labeyrie \cite{l70}). The
reconstructed visibilities and diffraction-limited images are
presented in Fig.\,\ref{f1}.

The results of two-dimensional elliptical uniform-disk fits to the
visibilities are summarized in Table~\ref{t2}. Our visibilities
are in very good agreement with the visibilities presented by
Danchi et al. (\cite{dtm01}). However, the diameters of the
best-fitting elliptical uniform-disk models presented in
Table~\ref{t2} are different from those derived by Danchi et al.
(\cite{dtm01}). This difference is not surprising because the
elliptical uniform-disk model does not match well the shape of the
observed visibilities, and hence a different fit range (due to the
different telescope diameters) results in different disk
diameters. For example, larger diameters ($J$: 78\,mas, $H$:
77\,mas, $K$: 79\,mas) are obtained if a fit range of up to only 6
cycles/arcsec is chosen.

The shape of the observed visibilities is more complex than that
of a simple Gaussian or uniform-disk model. At spatial frequencies
of $\sim$6 cycles/arcsec the steepness of all visibility curves
decreases considerably. This structure suggests a two-component
visibility model. Fig.~\ref{f2} shows best-fitting two-component
models, consisting of two Gaussian functions, fitted to the cuts
through the short and long axes of the two-dimensional
visibilities of MWC\,349A. Table~\ref{t3} summarizes the results
of all two-component fits (short and long visibility axes). The
visibilities along the short axis up to $\sim$6 cycles/arcsec
correspond to an extended object with a FWHM Gaussian fit diameter
of 94--119\,mas, whereas the visibilities (along the short axis)
at higher spatial frequencies correspond to the second, more
compact component with an angular diameter of only 22--27\,mas.
This two-component structure can be recognized because of the
small errors of the visibilities. These errors are small because
the speckle interferograms of the object (MWC\,349A) and of the
reference star (MWC\,349B), at a separation of only 2.4\,arcsec,
were recorded simultaneously. This has the advantage that the
object and the reference star speckle interferograms were recorded
under identical seeing conditions and therefore calibration errors
of the speckle transfer function are avoided.

An interpretation of the IR images of MWC\,349A in terms of a CS
disk seen nearly edge-on has been presented in detail by Danchi et
al. (\cite{dtm01}). They have shown that the elongated IR source
resolved in MWC\,349A can be fitted with a viscous accretion disk,
which possibly has a hole inside and is illuminated by a hot
central star. However, a viscous excretion disk also meets these
criteria (e.g., Lee et al. \cite{lso91} and Morris \cite{m81}). It
is therefore impossible to distinguish unambiguously between these
two models without considering the nature of the central source.
This analysis is presented in Sect.\,\ref{nature}.

   \section{Possible nature of MWC\,349A}\label{nature}

Three alternative points of view on the nature of MWC\,349A and,
correspondingly, three different classifications of the object
have been considered:
\begin{enumerate}
\item a Herbig\,Be star accreting material from a disk
(Thompson et al. \cite{thom77})
\item an evolved B[e] supergiant surrounded by an excretion disk
(Hartmann et al. \cite{hjh80})
\item a young planetary nebula
(Ciatti, D$^{\prime}$Odorico, \& Mammano \cite{com74})
\end{enumerate}

All these hypotheses are based on similarities between some of the
observational features of MWC\,349A and those of the corresponding
types of objects. In order to make the right choice, one first
needs to constrain the physical parameters of the stellar source
illuminating the CS material, since these parameters are the basis
for calculations of the physical conditions in the CS
environments. Let us review the existing information about the
underlying stellar source and summarize the constraints on its
parameters.

The effective temperature ($T_{\rm eff} \sim 35000$\,K) of
MWC\,349A was estimated by means of the Zanstra method, assuming
that the CS dust is external to the gas (Hartmann et al.
\cite{hjh80}), and from the absence of He\,{\sc ii} emission lines
in the spectrum. The latter finding is usually credited to Felli
et al. (\cite{f85}), who used the LTE Kurucz (\cite{k79}) model
atmospheres to calculate the He\,{\sc i} photoionization. However,
later Schmutz et al. (\cite{sch91}), in their study of an Ofpe/WN9
star \object{R84} with a weak He\,{\sc ii} line emission, showed
the importance of a NLTE approach and derived an upper limit of
T$_{\rm eff} \sim$28000\,K for the presence of He\,{\sc ii} lines
in emission. On the other hand, strong He\,{\sc i} emission lines,
such as those observed in the spectrum of MWC\,349A, are usually
seen only in emission-line objects with spectral types earlier
than B2, constraining a lower limit for the T$_{\rm eff}$ at
$\sim$20000\,K (e.g., Miroshnichenko et al. \cite{mir98a}). Thus,
the T$_{\rm eff}$ of MWC\,349A more likely corresponds to a
spectral type of B0--B1. The same conclusion has also been drawn
for B[e] objects with optical spectra similar to those of
MWC\,349A (e.g., \object{CI Cam}, Hynes et al. \cite{hyn02}). We
should note here that UV observations are needed to better
constrain T$_{\rm eff}$ for such a hot object. So far it seems
that only one observation of MWC\,349A has been obtained in this
spectral range: an IUE large-aperture short-wavelength spectrum,
which is rather noisy and contains a contribution from MWC\,349B.

In order to estimate the luminosity of MWC\,349A, one needs values
of its optical brightness, interstellar (IS) and CS extinction,
and distance. From the spectrophotometric data obtained in 1976
and 1979, C85 derived the continuum $V$=13.96\,mag. At that time
MWC\,349A, which was found significantly variable ($\Delta R
\sim$1.5\,mag, Jorgenson, Kogan, \& Strelnitski \cite{jks00}), had
nearly the mean brightness level.

The most difficult task is the extinction determination. Since no
photospheric lines are seen in the spectrum, a certain amount of
the CS continuum due to free-free and free-bound radiation
distorts the star's spectral energy distribution (SED). It is
unclear if the CS dust affects the SED, because the CS disk plane,
where the dust seems to be located, is inclined $\sim 20$\,
degrees to the line of sight. Therefore, if the disk is flat or
mildly flared, the dust may not contribute to the CS extinction.
C85 estimated the IS extinction to be between
A$_V$=8.8$\pm$0.1\,mag to 9.9$\pm$0.4\,mag, depending on the
assumption of whether the object's SED is affected by the CS
matter alone or not. This approach, however, provides no
information about the above mentioned CS contribution to the
object's brightness. At the same time, both mentioned A$_V$
estimates give effectively the same result for the extinction-free
brightness. The smaller A$_V$ corresponds to the accretion disk
veiling, which may well be of the order of 1\,mag, i.e. roughly
the difference with the other A$_V$ estimate, based on the normal
star SED. Thus, we can safely adopt a mean brightness of
$V$=14\,mag and A$_V$=9.9\,mag for MWC\,349A. More precise
estimates require a self-consistent modeling of both the continuum
SED and the emission-line spectrum.

There are two different estimates of the distance to MWC\,349A.
One assumes that the object belongs to Cyg\,OB2, whose distance is
1.7\,kpc (Kn\"odlseder \cite{knodl00}), while the other one is
based on the physical association of MWC\,349 A and B and an
assumption that MWC\,349B is a normal B0\,{\sc iii} star
(1.2\,kpc, C85). According to Neckel \& Klare (\cite{nk80}), the
IS extinction in the direction of MWC\,349 increases rapidly with
distance, reaching 4--5\,mag at $D$=1--1.5\,kpc. Analysing the
2MASS near-IR survey data, Kn\"odlseder (\cite{knodl00}) found
that for the objects belonging to Cyg\,OB2 (a region centered at
R.A.=20$^{h}33^{m}10^{s}$ and  Dec.=+41$^{\degr}12^{\prime}$
(J2000) with a radius of 1.05 degrees, which includes the position
of MWC\,349) the IS extinction varies from 5 to 20\,mag.
Altogether, these facts suggest that MWC\,349 is most likely a
member of Cyg\,OB2 at $D$=1.7\,kpc.

Combining the adopted values, one can estimate the absolute
magnitude of MWC\,349A to be M$_{V}=-7\pm1$\,mag (close to that
adopted by Hartmann et al. \cite{hjh80} after rescaling for the
distance difference, $-$6.7\,mag). From the above result for the
T$_{\rm eff}$, a bolometric correction of BC=$-2.5\pm$0.4\,mag
(Miroshnichenko \cite{mir98b}) can be deduced. Considering all
possible sources of uncertainty ($V$-magnitude, A$_V$, distance,
and BC), we derive a luminosity of $\log$ L/L$_{\sun}=5.7\pm$1.0.
The luminosity value is the most crucial parameter for
distinguishing between the possible models for this puzzling
object. The resulting error is assumed to be rather large, but
nevertheless of no influence on our further analysis. In the next
sections we consider all the hypotheses about the nature and
evolutionary state of MWC\,349A mentioned above.

\subsection{Pre-main-sequence star}\label{pms}

According to the theory of pre-main-sequence evolution, only stars
with initial masses of $\le$8--10 M$_{\sun}$ can be observed in
the optical region before they enter the main-sequence
evolutionary phase (Palla \& Stahler \cite{ps93}). The mass of
MWC\,349A, estimated by Ponomarev, Smith, \& Strelnitski
(\cite{pss94}) from the Keplerian velocities of the CS gas
(M$_*$=26\,(D/1.2 kpc)\, M$_{\sun}$), is out of the above range by
far. Also the object's luminosity, estimated above, is at least an
order of magnitude higher than those observed in Herbig Be stars
and those calculated by Palla \& Stahler (\cite{ps93}) for
pre-main-sequence models, even with the highest protostellar
accretion rate of 10$^{-4}$ M$_{\sun}$\,yr$^{-1}$. Furthermore, a
few known pre-main-sequence Herbig Be stars of early-B subtypes
(such as \object{MWC 1080} and \object{R Mon}) display much
stronger far-IR excesses, due to the radiation of the cold dust
from the protostellar cloud, than that observed in MWC\,349A. The
IRAS and ISO (Thum et al. \cite{th98}) fluxes decrease rather
steeply longward of $\sim 25\mu$m, indicating a lack of cold dust
in the immediate surroundings of the object. The presence of the
secondary component in pre-main-sequence stars does not affect the
outer regions of the CS envelope, since almost all early B-type
Herbig Be stars are known to be binaries (e.g., Leinert, Richichi,
\& Haas \cite{lrh97}).

In principle, the cold dust may be destroyed by a strong diffuse
UV radiation field of hot stars in Cyg\,OB2, which contains more
than 100 O-type stars and is considered to be a young globular
cluster (Kn\"odlseder \cite{knodl00}). However, if MWC\,349A does
belong to Cyg\,OB2, it is hardly a pre-main-sequence object,
because many cluster members are evidently more evolved (a few WR
stars and LBV candidates, as well as $\sim$100 O-type stars, are
known in Cyg\,OB2) and seem to have been born at roughly the same
time. It also seems unlikely that MWC\,349A is much closer to us,
because different data, such as photometry and spectroscopy (C85)
as well as polarimetry (Meyer et al. \cite{mnh02}), indicate the
presence of a large IS reddening inconsistent with a distance less
than 1\,kpc.

Another characteristic, which needs to be considered here, is the
shape of the object's spectrum in the 10\,$\mu$m region. The
features seen in this region are indicative of the chemical
composition and/or the optical depth of the CS dust. The spectrum
of MWC\,349A is essentially featureless (except for the line
emission) between 8 and 13\,$\mu$m (e.g. Rinehart, Houk, \& Smith
\cite{rhs99}). This suggests either that the optical depth of the
dust is large or that the CS dust is mostly composed of amorphous
carbon. The latter is not consistent with the very young age of
the object, as the carbon-rich dust is usually observed around
evolved objects, which create the dust from the material processed
in the stellar interiors. From all of the above, we can conclude
that MWC\,349A is unlikely to be a pre-main-sequence object.

\subsection{Young planetary nebula}\label{ypn}

Some of the observed features of MWC\,349A are similar to those of
post-AGB stars, which evolve toward the planetary nebula phase
(for example, the compact nebulae). However, such stars produce
dust only during the AGB phase. Later the dust moves slowly
further away from the star, becomes colder, and radiates only in
the far-IR region of the spectrum. This is not the case for
MWC\,349A. Additionally, the derived luminosity of the object
exceeds those of the most massive proto-planetary nebulae by at
least an order of magnitude (e.g. Bl\"ocker \cite{bl95}). Thus, it
seems possible to rule out the suggestion that MWC\,349A is a
post-AGB object.

\subsection{B[e] supergiant}\label{besg}

The adopted T$_{\rm eff}$ and luminosity of MWC\,349A ($\log$
L/L$_{\sun} \approx 5.7\pm1.0$), along with the strong
emission-line spectrum and the SED shape, make it very similar to
B[e] supergiants, a type of evolved star first discovered in the
Magellanic Clouds (Zickgraf et al. \cite{z86}). Their properties
include the co-existence of broad emission lines of
high-excitation species (e.g. C\,{\sc iv}) with P\,Cyg-type
profiles in the UV region and much stronger single- or
double-peaked line profiles of low-excitation species in the
optical region (e.g. H\,{\sc i}, He\,{\sc i}, and Fe\,{\sc ii}).
Zickgraf et al. (\cite{z86}) explained such behaviour using a
model in which the UV lines are formed in the fast and hot stellar
wind in the polar regions of the CS envelope, while the optical
lines are formed in the slower, denser, and cooler wind, forming
the excretion disk. The SEDs of the B[e] supergiants also seem to
be similar to that of MWC\,349A. Their IR fluxes decrease towards
longer wavelengths and no dusty features are seen in the
10\,$\mu$m region. Bjorkman (\cite{bj98}) modelled the SED of
\object{R 126}, a typical B[e] supergiant from the LMC, and showed
that the CS dust (which was considered to be composed from mostly
amorphous carbon) can be formed in its disk. From these
similarities and the parameters we adopted for MWC\,349A, it seems
more likely that the object is a B[e] supergiant, rather than a
pre-main-sequence or a post-AGB star.

\section{Discussion}\label{discuss}

Our observations, presented in Sect.\,\ref{obsres}, suggest that
the IR source surrounding the central star of MWC\,349A consists
of two components. For example, in the $H$-band the diameter
(along the major axis) of the Large Component is $115.7\pm
7.6$\,mas, corresponding to $190\pm 13$\,AU (for a distance of
$D$=1.7 kpc), while the diameter of the Small Component is only
$26.9\pm 3.0$\,mas, corresponding to $41\pm 4$\,AU.

One of the possible interpretations of this finding is that the
emission of the inner parts of the disk is dominated by the
bound-free and free-free radiation of the CS gas, while the outer
parts of the disk are observed mainly due to emission of the CS
dust. This interpretation is supported by the strong emission-line
spectrum of MWC\,349A, which implies a significant contribution of
the CS gas continuum emission in the red and near-IR spectral
region (up to $\lambda \sim$ 1--2\,$\mu$m). Moreover, the CS gas
would occupy a less extended spatial region -- closer to the star
-- than the dust. Therefore, the CS gas can be observed as an
enhancement in the visibilities at high spatial frequencies.
Within this interpretation the size of the more compact component
indicates the radial distance at which the dust sublimation in the
disk occurs.

We can also consider another interpretation of the observed
two-component structure. The structure of the IR source in
MWC\,349A could be caused by a discontinuity of the mass transfer
rate in the CS disk. If this discontinuity does occur, a
significant fraction of the kinetic energy of the material flowing
in the disk will be converted into thermal energy and radiation.
Hence, within this hypothesis an additional source of radiation,
localized at the outer edge of the compact component, is expected.
Such a source can be associated with the $H30\alpha$ recombination
line maser, which according to Planesas, Mart\'in-Pintado, \&
Serabin (\cite{pmps92}) is generated in the CS disk at a radial
distance of about 40\,AU from the central star of MWC\,349A.
Though this association remains to be justified, it suggests that
a more detailed modeling of the mass transfer discontinuity might
be very fruitful.

Among the possible reasons for the discontinuity, one can indicate
an instability of the CS disk and the interaction between the
material presently ejecting from the central star with a fossil
disk-like shell, which was formed during a previous epoch of
evolution of MWC\,349A. The existence of a fossil shell is not
very unusual, if MWC\,349A is an evolved star (see
Sect.\,\ref{nature}). Furthermore, it cannot be excluded that
MWC\,349A is a close binary system. Indeed, C85 admitted that
their L and M$_V$ estimates do not fit those of any normal star.
They suggested that the problem lies either in the unusually slow
and dense wind or in the presence of a cool companion (also
proposed earlier by Blanco \& Tarafdar \cite{bt78}), whose
spectral signature however has not been detected. Another type of
the secondary (a massive protoplanet) was suggested by Jorgenson
et al. (\cite{jks00}), who found regular variations in the
object's optical brightness with a period of 9 years, but stated
that the brightness variation mechanism is unclear. This idea is
also incompatible with our conclusion about the evolutionary state
of MWC\,349A. At the same time, the binary hypothesis is worth
additional consideration, although we cannot rule out that
MWC\,349A is a single B[e] supergiant. The presence of a secondary
companion makes it easier to explain such properties of MWC\,349A
as the extremely strong Balmer emission and a non-spherical
distribution of the CS matter. We note here that MWC\,349B is not
considered physically connected with MWC\,349A.

A few B[e] supergiants have been found in the Milky Way. The
observed features of one of them (CI Cam, e.g. the emission line
profiles and the SED shape) are for the most part very similar to
those of MWC\,349A. There have been no attempts to resolve the
envelope of CI Cam, which seems to be located further away from
the Sun (at 3--5\,kpc, see Robinson, Ivans, \& Welsh \cite{riw02}
and Miroshnichenko et al. \cite{mir02} for different points of
view), with speckle interferometry. This object experienced a
strong outburst, detected in all spectral regions from the
$\gamma$-ray to the radio, which was explained by the interaction
in a binary system, containing a B[e] supergiant and a compact
degenerate companion (a neutron star or a black hole, see Robinson
et al. \cite{riw02}). It is difficult to detect secondary
companions in systems where it is invisible in the quiescent
state. The same problem concerns the B[e] supergiants of the
Magellanic Clouds, as they are faint and very distant. However,
Zickgraf et~al. (\cite{zk96}) reported that one of these objects,
\object{R4}, has been proven to be a binary through the detection
of regular radial velocity variations. Thus, binarity may be a key
property to explain the observational features of B[e]
supergiants. However, such systematic observations as spectroscopy
and optical photometry (not available yet for MWC\,349A) are
needed in order to find traces of the secondary and put
constraints on its properties.

The 9-year period found by Jorgenson et al. (\cite{jks00}) may be
due to orbital motion. Assuming a system mass of 26\,M$_{\sun}$,
these authors mentioned that such an orbital period would
correspond to a separation of 12--15\,AU, which in turn
corresponds to an angular separation of $\sim$7--10\,mas at a
distance of 1.7\,kpc. To verify this, an order of magnitude of
better spatial resolution than that of the existing observations
is needed. Such a resolution can soon be achieved with modern
interferometers.

\section{Conclusions}\label{conclus}

We have obtained and analyzed diffraction-limited (resolutions
43--74\,mas) speckle interferometric $J-$, $H-$, and $K$-band
observations of MWC\,349A. The images are elongated at all
wavelengths with an axial ratio of $\sim$0.7. Their position
angles coincide with those obtained by Danchi et al. \cite{dtm01}.
The best-fit model to the observed visibilities consists of 2
elliptical components (e.g. $J$-band size: $\sim 22\times12$\,mas
and $\sim 119\times74$\,mas; Table\,\ref{t3}). The inner component
probably corresponds to the CS gas emission, while the outer one
can be produced by a dusty disk, whose symmetry plane is seen at a
low inclination angle.

We also summarized existing information about observed parameters
of MWC\,349A, and hypotheses previously suggested to explain its
nature. This analysis resulted in a new luminosity estimate, which
is based on the assumption that the object belongs to Cyg\,OB2
(see Meyer et~al. \cite{mnh02} for the justification) and
indicates that it is more luminous than was suggested before. We
concluded that MWC\,349A is unlikely to be a pre-main-sequence
star or a young planetary nebula. Its fundamental parameters and
observed features (the strong emission-line spectrum and dust
emission) are more consistent with those of B[e] supergiants.

A binary nature of MWC\,349A cannot be excluded, but this
hypothesis needs further observational testing. In particular,
optical photometric observations to search for possible brightness
variations, and high-resolution spectroscopy to search for radial
velocity variations both associated with the orbital motion are
very important for further progress in the understanding of this
puzzling object. Quantitative analysis of our speckle images
within this model will be presented elsewhere.

\begin{acknowledgements}
We would like to thank Peter Tuthill and the anonymous referee for
useful comments. Nazar Ikhsanov acknowledges the support of the
Alexander von Humboldt Foundation within the Long-term Cooperation
Program.

\end{acknowledgements}

\listofobjects
\end{document}